\documentclass[acmsmall,authorversion,nonacm]{acmart}   

\AtBeginDocument{%
  \providecommand\BibTeX{{%
    \normalfont B\kern-0.5em{\scshape i\kern-0.25em b}\kern-0.8em\TeX}}}

\usepackage{url}
\usepackage{doi}

\usepackage{currfile}
\usepackage{textcomp}

\usepackage{enumitem}

\begin{document}


\title{Blockchain Gateways, Bridges and Delegated Hash-Locks}

\author{Thomas Hardjono}
\affiliation{%
\institution{MIT Connection Science}
\streetaddress{77 Massachusetts Avenue}
\city{Cambridge, MA 02139}
}
\email{hardjono@mit.edu}



\begin{abstract}
In the current work we discuss the notion of gateways as a means
for interoperability across different blockchain systems.
We discuss two key principles for the design of gateway nodes and scalable gateway protocols,
namely (i) the {\em opaque ledgers principle} as the analogue of the autonomous systems principle
in IP datagram routing, and 
(ii) the {\em externalization of value principle} as the analogue of the
end-to-end principle in the Internet architecture.
We illustrate the need for a standard gateway protocol
by describing a unidirectional asset movement protocol between two peer gateways,
under the strict condition of both blockchains being private/permissioned
with their ledgers inaccessible to external entities.
Several aspects of gateways and the gateway protocol is discussed,
including gateway identities,
gateway certificates and certificate hierarchies,
passive locking transactions by gateways,
and the potential use of delegated hash-locks to expand the functionality of gateways.\\
~~\\
{February 11, 2021}
\end{abstract}





\maketitle

{\small 
\tableofcontents
}

\newpage
\clearpage


\section{Introduction}
\label{sec:Introduction}

Interoperability across blockchain systems represents one of the open challenges
today for the broad adoption of blockchain technology and distributed ledger technology (DLT) generally.
The promise of the decentralization of control as offered by the vision of blockchains
cannot be attained if different blockchain systems 
-- employing different internal ledger architectures, consensus paradigms and differing
processing incentives -- cannot interoperate at the scale of the Internet.
In the current work we explore further the notion of {\em blockchain gateways}
as a means to interconnect distinct and incompatible blockchain networks~\cite{HardjonoLipton-IEEETEMS-2019}.

In Section~\ref{sec:Interoperability} we motivate the need for gateways
from the perspective of the autonomy of blockchain networks
and the need to address the problem of cross-border virtual asset movements~\cite{FATF-Recommendation15-2018}.
Section~\ref{sec:GatewayArchitecture}
proposes two key principles for the design of gateway nodes and scalable gateway protocols.
In Section~\ref{subsec:2PC-protocol} we illustrate with a gateway protocol
that performs a unidirectional movement of a virtual asset representation from one blockchain to another. 
The various aspects and open challenges to the proper design of blockchain gateways
are then discussed in Section~\ref{sec:Dicussion}.
We also explore additional possible functions that a gateway may implement in Section~\ref{sec:FurtherAfield},
including the use of a delegated hash-lock to a remote third party blockchain
to assist in the movement of a fungible virtual asset and of the value that may be associated with the asset.

We seek to make this paper readable to a broad audience,
and thus have avoided using detailed technical constructs belonging specific blockchains.
The reader is assumed to have some basic knowledge
regarding blockchains, shared ledgers,
consensus algorithms and public-key technology.
We use the term ``blockchain'' and ``DLT'' interchangeably for convenience.

\section{Gateways for Interoperability, Autonomy and Asset Mobility}
\label{sec:Interoperability}

Currently blockchains and DLT systems
are exhibiting similar patterns of growing pains as that of LAN technologies in the 1980s.
There is today a proliferation in the number of blockchain projects
around the world (e.g. over three dozen systems as surveyed by~\cite{Belchior2020-survey}).
Although on the surface this apparent proliferation of blockchain projects
may seem to be a new phenomenon,
in reality this is part of the natural cycle in the development of emerging technologies.
A similar development cycle was also experienced
in the early days of Local Area Networks (LAN) technologies
and the routing protocols for data within these LANs.
Examples of Enterprise LAN technologies
from that period of development include IBM~SNA, DECnet, Ethernet, etc.

One milestone moment in the history of the nascent Internet was the definition
of the IP datagram structure~\cite{CerfKahn74}
with the support of DARPA and their goals for the survivable network~\cite{Clark88,Clark2018-book}.
The architecture of the Internet viewed each network as an {\em autonomous system} (AS),
where each AS would operate its own interior routing protocol.
A given autonomous system may even contain multiple LANs,
where each LAN could implement different physical layer (PHY level) protocols
(e.g. token-ring in SNA, CSMA/CD in DECnet, etc).
This approach permitted each autonomous system to develop and deploy its own technological choice
to satisfy the goal of delivering the standard IP datagram,
from the point where the IP datagram enters the autonomous system
to the point where it departs from it.
Each autonomous system was free to operate the interior (intra-domain) 
routing protocol of its choice (e.g. RIP, IS-IS, OSPF, etc,)~\cite{Perlman1999}.
In order to interconnect autonomous systems to provide end-to-end reliable services to the user,
special routers called {\em border gateway} routers
were deployed between peered autonomous systems.
The main function of a border router in a given autonomous system
is to advertise available routes to its peered border router at an adjacent autonomous system.
Protocols such as the Border Gateway Protocol (BGPv4)~\cite{RFC1105-formatted}
were developed for this purpose.

In order for blockchain-based services to scale globally,
blockchain networks must be able to interoperate with one another
following a standardized protocol and interfaces (APIs).
We believe that a {\em blockchain gateway} is needed for blockchain networks
to interoperate
in a manner similar to border gateway routers in IP networks.
Furthermore,
just as border gateway routers use the BGPv4 protocol to interact with one another
in a peered fashion,
we believe that a {\em blockchain gateway protocol}
will be needed to permit the movement of virtual assets and related information
across blockchain networks in a secure and privacy-preserving manner.
We motivate the need for blockchain gateways and blockchain gateway protocols
in the following summary:
\begin{itemize}[topsep=8pt,itemsep=1pt, partopsep=4pt, parsep=4pt]

\item	{\em Enables blockchain interoperability}: 
Blockchain gateways provide an interface for the interoperability
between blockchain/DLT systems that operate distinct consensus protocols
and ledger data structures. 
A gateway fronts its blockchain/DLT system,
and exposes a standard interface (APIs) to an external peer (opposite) gateway.
Facing outwards,
a gateway implements a gateway-to-gateway protocol with the peer gateway.
Facing inwards,
a gateway implements the relevant interior resources 
(e.g. consensus protocol, ledger management, key management, etc.)
relevant to participate
in its blockchain/DLT system.
Gateways permits private blockchain/DLT systems
to interact with public blockchain/DLT systems.

\item	{\em Ensures blockchain network autonomy}:
The use of gateways as the interface point between blockchain networks
permits each blockchain network to evolve,
where new innovations and new technologies can be deployed interiorly
within a blockchain network without affecting other external blockchains.
In this way,
a blockchain network truly behaves as an {\em autonomous network} in
the same vein as the original vision of the IP Internet.

\item	{\em Enables virtual asset mobility}: 
There is a growing need for virtual assets to be moveable across blockchain networks,
a need that will only increase with growth of CBDC tokenized currencies~\cite{AuerCornelli2020-CBDC}.
The use of blockchains permit innovative {\em asset movement protocols}
to be developed that can be implemented by gateways across standardized APIs.
Such asset movement protocols can be designed for specific asset types
and be operated by gateways
according to the different regulatory jurisdictions in the world.

\item	{\em Enforcement point for AML/KYC regulations and international taxation}: 
Gateways as the ``landing points'' for virtual asset entering into (departing from)
blockchain networks becomes an enforcement point
for AML/KYC regulations~\cite{FATF-Recommendation15-2018}.
Furthermore,
for cross-border movement of assets the gateways also become
``checkpoints'' where international taxation concerns can be addressed and implemented.

\item	{\em Enables integration with legacy systems}:
The use of a standardized gateway-protocol between peers of gateways
permits one of the blockchains to be substituted for a legacy system
(e.g. financial database systems)
without impact to the remote peer.
That is, a standardized gateway protocol can be use to hide the fact behind
the gateway lies a legacy system.
The gateway hides the complexity of the interiors of the system that it fronts
-- be it legacy data systems or new blockchain/DLT systems.

\end{itemize}

\begin{figure}[t]
\centering
\includegraphics[width=0.9\textwidth, trim={0.0cm 0.0cm 0.0cm 0.0cm}, clip]{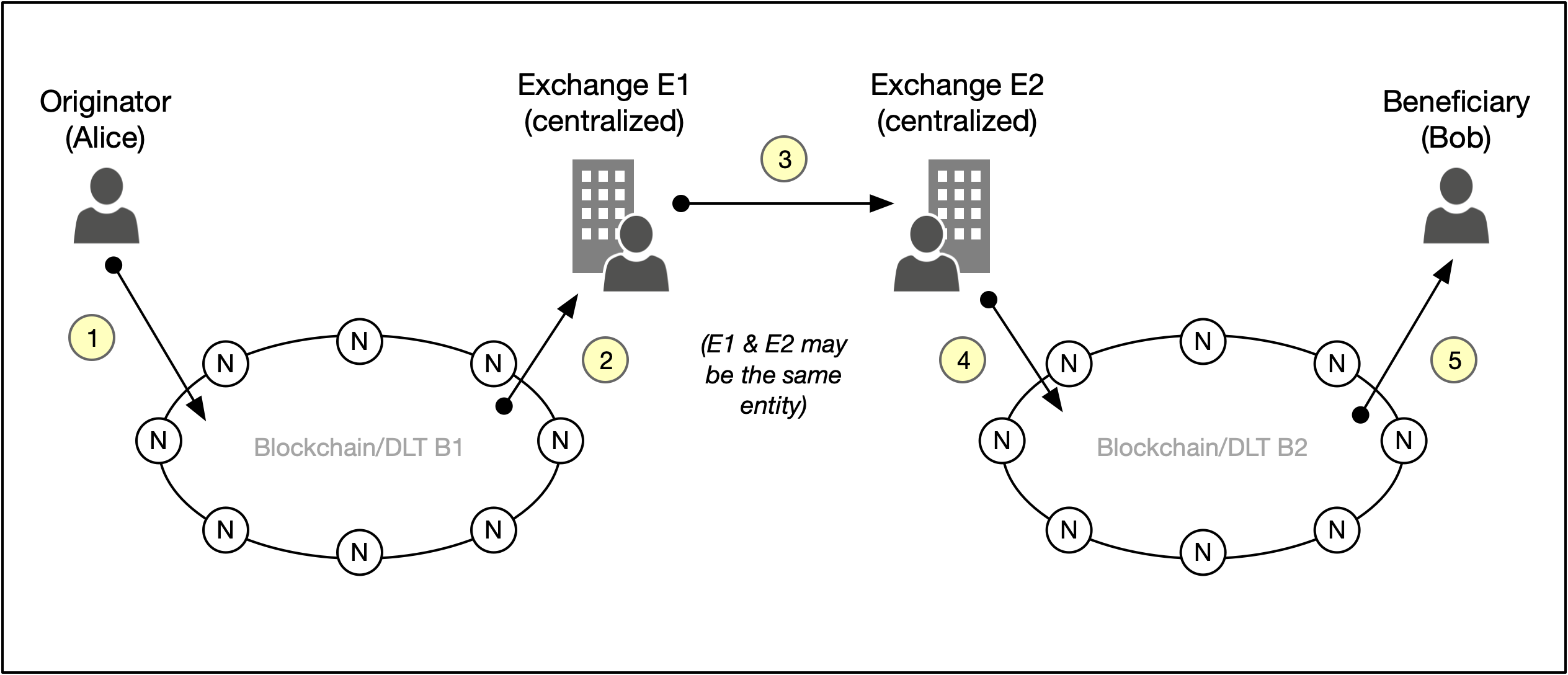}
\caption{The problem of dependence on centralized exchanges for asset movement}
\label{fig:gateway-bridges}
\end{figure}

\section{Blockchain Gateway Architecture}
\label{sec:GatewayArchitecture}

In this section we discuss the concept of gateways in the context of asset movement
across blockchain networks.
Our view is motivated by the fact that:
(i) there are today and will increasingly be multiple autonomous blockchain networks in the future;
(ii) that many of these blockchains will be private with ledgers that are read/write permissioned;
and that
(iii) interoperability among these blockchains (public and private)
will be a core necessity that will determine the viability of blockchain and DLT technology
as a means to retain and exchange digital representations of economic value.

The design of a gateway architecture has many challenges,
three among which are:
(i) {\em entity identities} -- namely the correct validation of 
the digital identities of entities in the blockchain ecosystem;
(ii) {\em atomic asset movement} -- namely the secure, atomic and efficient movement of the digital 
representation of virtual assets from one blockchain to another;
and
(iii) {\em reliability} of gateways and the mechanisms to recover securely from crashes.

\subsection{Gateway Definition, Roles and Functions}
\label{subsec:GatewayDefinition}

A {\em blockchain gateway} is a computer system in a blockchain network
for the purpose of assisting in the movement of virtual assets
into (out of) the blockchain network~\cite{HardjonoLipton-IEEETEMS-2019,HardjonoLipton2020-BookChapter}.
Being a node in its blockchain network, 
a gateway has read/write 
access to the shared ledger of the blockchain
and may participate in the consensus mechanism of the blockchain.
A gateway is dual-faced in that it is said to be {\em facing interiorly} when interacting
with its blockchain,
and {\em facing exteriorly} when interacting with a remote {\em peer gateway} 
belonging to a different blockchain network.
Two (2) peer gateways jointly execute a {\em gateway protocol} that implement
the relevant steps for moving a virtual asset with the mediation of the gateways
from one blockchain to another.

By definition a gateway belongs to one blockchain network only,
and when facing interiorly (inward)
it must be identifiable and authenticable to entities in that blockchain network.
For interior interactions,
a gateway has at least one blockchain {\em transaction signing} public/private key-pair.
When facing exteriorly,
a gateway must have a {\em gateway identity} public/private key-pair
that it employs when interacting with remote peer gateways.
This key-pair permits the gateway to be 
identifiable and authenticable by peer gateways
(see Section~\ref{subsec:GatewayIdentitiesKeys} below for a discussion on gateway identity keys).

The legal owner and operator of a node
is referred to as a {\em Virtual Asset Service Provider} (VASP) in the sense
of the FATF definition~\cite{FATF-Recommendation15-2018}.
The VASP owner of a gateway employs
a {\em VASP identity} public/private key-pair.
Some form of cryptographic binding is assumed
to exists between the gateway identity public-key
and the VASP identity public-key.

\begin{figure}[t]
\centering
\includegraphics[width=1.0\textwidth, trim={0.0cm 0.0cm 0.0cm 0.0cm}, clip]{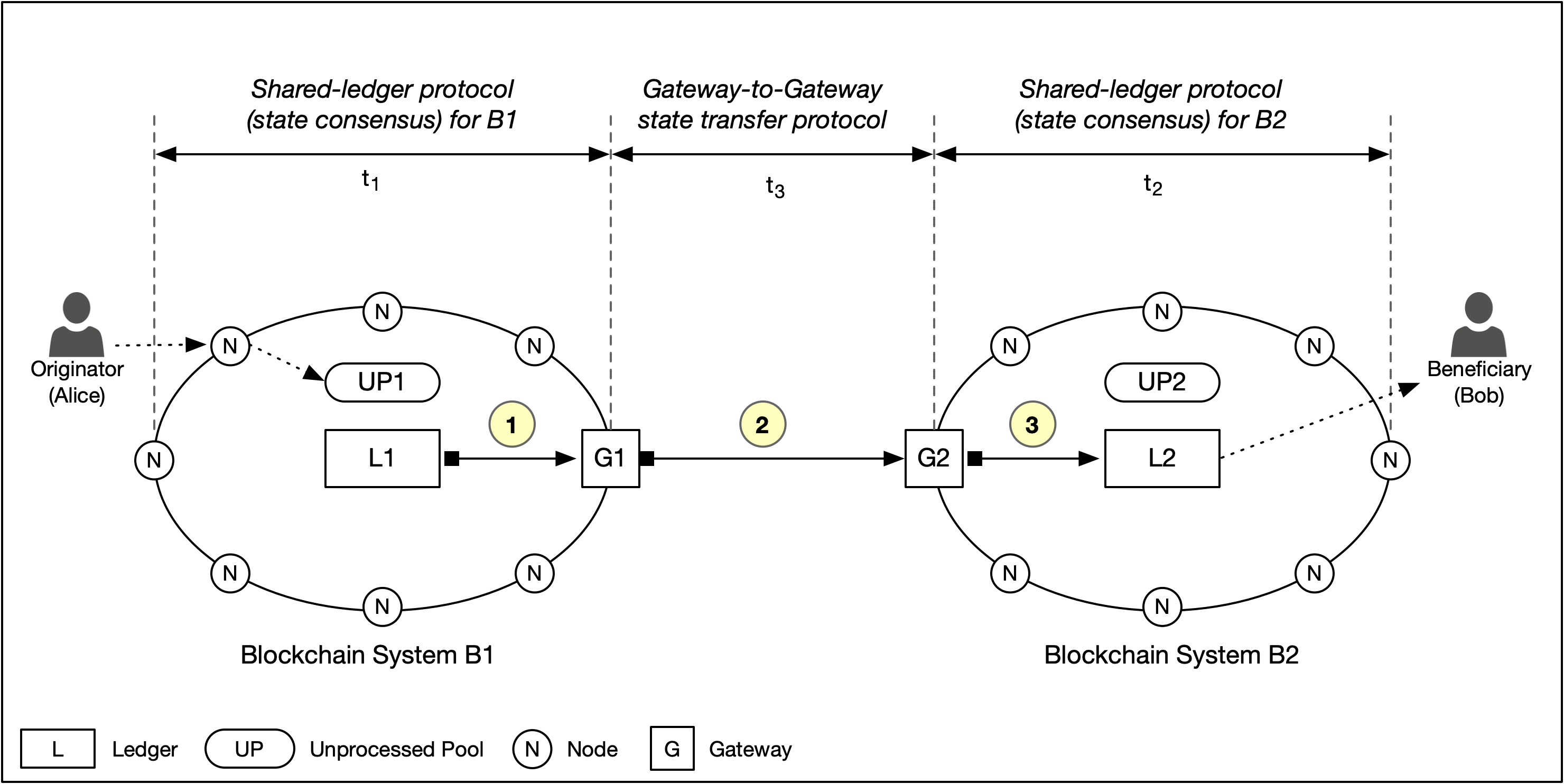}
\caption{Three (3) segments of the cross-chain asset movement problem}
\label{fig:ThreeSegments}
\end{figure}

At a high level,
among others there are three (3) roles and functions of peer gateways (Figure~\ref{fig:ThreeSegments}):
\begin{itemize}[topsep=8pt,itemsep=1pt, partopsep=4pt, parsep=4pt]

\item	{\em Protocol and state information translation}:
A blockchain gateway performs the translation of ledger-based state information
from one blockchain to another.
The state information in a ledger pertains to the ownership of a virtual asset,
which typically is associated with (bound to) a public-key (address) of the current owner.

A ledger within an origin blockchain may employ a different data structure
(to represent virtual assets) that is different from the 
ledger data structure used in the destination blockchain.
As such, the gateway in an origin blockchain
may need to deliver the information in a standardized format
to the gateway in an destination blockchain
in order to permit interoperability.
Looking at Figure~\ref{fig:ThreeSegments},
gateway G1 in blockchain B1 may translate the information
from its ledger L1,
while gateway G2 in blockchain B2 performs the same task for its ledger L2.
The gateway protocol employed between gateway G1 and G2
should define a standardized format
that can convey acurrately
the information from ledger L1 to L2.

\item	{\em Ledger consistency maintenance during asset movement}:
Individually, a gateway maintains the consistency of the ledger in its blockchain system
in connection to the movement of virtual assets into (out of) the blockchain system.
Together, two peer gateways must jointly
maintain consistency of state information across their ledgers respectively.

In the context of virtual asset movement assisted by G1 and G2,
{\em consistency across ledgers} L1 and L2 
refers to the prevention by G1 and G2 of the double-existence of 
the asset (leading to double spending) in blockchains B1 and B2 respectively.

\item	{\em Regulatory Policy Enforcement Point}: 
A gateway takes on the role of a Policy Enforcement Point (PEP)~\cite{RFC2753-formatted-AAA,NIST800-207-ZeroTrust}
for enforcing rules (policies) that implement regulations
related to asset types and legal jurisdictions.

The movement of a virtual asset between blockchains 
may in fact cross regulatory and governance boundaries,
and as such information regarding the legal status of the asset and the involved entities
may reside outside the ledger of the blockchain.
The function of a gateway is also to obtain validated information
from eternal sources regarding the asset and entities.

\end{itemize}

\subsection{The Three Parts of the Cross-Chain Asset Movement Problem}
\label{subsec:ThreeParts}

In the current work we subdivide the problem of a gateway protocol
for cross-blockchain asset movement
into three (3) parts or sub-problems,
where the three steps  
must execute to completion in atomic fashion (see Figure~\ref{fig:ThreeSegments}):
\begin{enumerate}[topsep=8pt,itemsep=1pt, partopsep=4pt, parsep=4pt]

\item	{\em Extinguishment of the asset at the origin blockchain}: 
Information must be recorded on the origin ledger L1
to denote the fact that the asset associated with its last owner (originator address)
no longer exists (i.e. no longer accessible)
in the origin blockchain.
This is shown as Step-1 in Figure~\ref{fig:ThreeSegments}.

\item	{\em Blockchain-to-blockchain transfer commitment}: The gateways that
mediate (i.e. delegated to perform) the asset movement
must exchange signed messages that legally commit
to performing the relevant steps of the asset-movement.
This is shown as Step-2 in Figure~\ref{fig:ThreeSegments}.

\item	{\em Regeneration of the asset at the destination blockchain}: Information 
must be recorded on the destination ledger L2 that an asset instance has been
introduced into the ledger and is associated with an owner (beneficiary address) in that blockchain.
This is shown as Step-3 in Figure~\ref{fig:ThreeSegments}.

\end{enumerate}

Since the ledger of a blockchains operates on an {\em append-only} basis~\cite{HaberStornetta1991,BayerHaber1993},
this means that unlike traditional databases
there is no way to ``delete'' information
that has been recorded (confirmed) on the ledger.
This means that in reality the task of ``extinguishment''
and ``regeneration'' of a virtual asset
must be implemented using the same append-only mechanism used by the blockchain.
In other words,
the same transaction processing paradigm used in blockchain technologies
should be used to manifest the notions of extinguishment and regeneration of assets.
We discuss further the possible mechanisms to ``mark'' ledgers 
for asset extinguishment
and regeneration in Section~\ref{subsec:ExtinguishmentRegeneration}.

It is important to note that a cross-blockchain asset movement
is not necessarily equivalent to a local movement (same blockchain).
It is insufficient to simply record on the ledger L1
that the asset's ownership was transferred to a given address (public key),
in this case to the address of the gateway G1.
This is due to the possibility that the destination blockchain B2
may reside within a different regulatory and governance jurisdiction.
Thus,
the transaction initiated by the originator on blockchain B1
must include both the address of the beneficiary and the blockchain identifier
where the beneficiary address is located.
This combined information -- including the public-keys of G1 and G2 --
must be recorded on both ledgers L1 and L2,
including when both blockchains B1 and B2 are private/permissioned.
This recording on ledgers L1 and L2 must be such that
there is little room for disputes (e.g. among the originator, beneficiary, owner of G1 and owner of G2)
and that post-event audits can be performed on both sides.

\subsection{Design Principles}
\label{subsec:DesignPrinciples}

The interoperability of blockchain systems affects the scalability of the services
built atop the blockchains.
In order to develop scalable gateway protocols,
there are a number of design principles underlie 
the interoperability architecture of gateways~\cite{HardjonoLipton2020-BookChapter}:

\begin{quote}
	{\bf \em Opaque ledgers principle}: The interior ledger and 
	other resources of each blockchain system 
	must be assumed to be opaque to (hidden from) external entities.	
	Any resources to be made accessible to an external entity 
	must be made explicitly accessible by its gateway with proper authorization.\\
\end{quote}
This principle is the analogue of the {\em autonomous systems principle} in IP networking~\cite{Clark88}, 
where interior routes in local subnets are not visible to other external autonomous systems.
Observance of this principle leads to a design of gateway-to-gateway protocols
that do not rely on (be dependent on) interior constructs
that are unique to specific blockchain systems.\\

\begin{quote}
	{\bf \em Externalization of value principle}: The gateway-to-gateway protocol 
	that implements the cross-blockchain asset movement
	must be agnostic to (oblivious to) 
	the economic or monetary value of the virtual asset being moved.\\
\end{quote}
This principle is the analogue of 
the {\em end-to-end principle} in the Internet architecture~\cite{SaltzerReed84}, 
where contextual information of the communicating parties
is placed at the endpoints of the communications instance.	
The Internet's interior routing fabric is oblivious as to which
application each datagram belongs to.
This principle permits gateway protocols to be 
designed for efficiency, speed and reliability -- independent of
the changes in the perceived economic value of the virtual asset.
In the case of virtual asset movements, the originator and beneficiary 
at the respective blockchain systems are assumed to have 
a common agreement regarding the economic value of the asset.

The resources within a blockchain system (e.g. ledgers, keys, consensus protocols, smart contracts, etc.) 
are assumed to be opaque to external entities in order to permit 
a resilient and scalable protocol design that is not dependent on 
the interior constructs of particular blockchain systems.	
This ensures that the virtual asset movement protocol between gateways 
is not conditioned upon or dependent on these interior technical constructs.	
A gateway interacts in two directions, namely facing outbound or exteriorly to other peer gateways
and facing inbound to its blockchain system.
The role of a gateway therefore is also to mask (hide) the complexity of the 
interior constructs of the blockchain system that it represents.	
Overall this approach ensures that a given blockchain system 
operates as a true autonomous system.

The point of the opaque ledgers/resources principle is to enable the design
of a cross-blockchain asset movement protocol
under the {\em strictest condition} -- namely that both blockchains
are private and their ledgers, keys, data structures and smart-contracts (i.e their interior resources)
are inaccessible to each other -- 
so that the same protocol can be used in less strict situations also.
Thus, the interaction between the two private blockchains
are possible only through their respective gateways
and the interfaces (i.e. standardized APIs) which the gateways implement.
The thinking is that if an asset movement protocol  (i.e. gateway protocol)
can operate securely and efficiently
between two private blockchain systems (with no visibility into each other's ledgers/resources),
then the protocol will also operate
when one or both blockchains are public/permissionless.

\subsection{Gateway Protocols: Desirable Atomicity Properties}
\label{subsec:DesirableAtomicityProperties}

From the state (ledger) consistency perspective,
there are a number of desirable properties
with regards to a gateway protocol~\cite{IETF-draft-hardjono-gateways}:
\begin{itemize}[topsep=8pt,itemsep=1pt, partopsep=4pt, parsep=4pt]

\item	{\em Atomicity}: An asset movement must either commit or entirely fail (where failure
means there are no changes to asset ownership in the origin blockchain).

\item	{\em Consistency}:  An asset movement (commit or fail) must always leave both
blockchain systems in a consistent state,
where the asset in question is located in one blockchain only.
A protocol failure must not result in a ``double-existence'' of the asset 
(leading to double-spend in the two blockchain systems respectively).

\item	{\em Isolation}: While the asset movement is occurring, 
the asset ownership cannot be modified.
That is, some kind of temporary disablement of the asset at the origin blockchain
must be used (e.g. locking on the ledger, escrow to a gateway, etc).
This is to prevent the owner (who requested the cross-chain asset movement)
from double-spending the asset locally while the gateway protocol is executing.

\item	{\em Durability}: Once an asset movement has been committed on both the
origin blockchain and destination blockchain, this commitment must remain true
regardless of subsequent gateway crashes.

\end{itemize}
These properties must hold true regardless of whether one or both of the blockchain systems
are private (permissioned) or public (permissionless).

\subsection{Gateway Protocols: Desirable Security Properties}
\label{subsec:DesirableSecurityProperties}

From a security and non-repudiation perspective,
there are a number of requirements for a gateway-protocol:
\begin{itemize}[topsep=8pt,itemsep=1pt, partopsep=4pt, parsep=4pt]

\item	{\em Identification and authentication}: 
Gateways must be able to mutually identify and authenticate each other
using the relevant gateway identity keys, trust anchors and other entity identifier schemes.
This means that the gateway identity key-pair used to interact externally
with other peer gateways 
must be different from the transaction signing key-pair that it
uses internally for its ledger.

\item	{\em Secure channel for confidentiality and privacy}: 
All communications between gateways
must be be performed over a secure channel,
using standard secure channel establishment protocols
(e.g. IPsec~\cite{RFC2401-formatted,RFC2409-formatted,RFC2408-formatted}, 
SSL/TLS~\cite{RFC8446-formatted,NIST800-52-TLS-Guide}).

\item	{\em Non-repudiation of asset movement}:
Relevant parts of the messages exchanged between gateways
must be digitally signed using the gateway identity key-pair
in such a way that the entire protocol exchange yields
non-repudiable evidence regarding the movement of the asset.
Gateways must keep a forensic log of these messages
in such a manner that the complete flows are reconstructable 
in the case of a future legal dispute or a future inquiry by relevant regulatory authorities.

\item	{\em Crash recovery into secure state}:
The crash of one gateway (or both) during the movement of an asset
must not result in loss or leakage of information
pertaining to asset or entity information (e.g. originator and beneficiary identities).
The gateway recovery mechanism must always place a gateway into a safe state.

\end{itemize}
We discuss gateway identities and keys in Section~\ref{subsec:GatewayIdentitiesKeys}.

\section{An Atomic Unidirectional Gateway Protocol}
\label{subsec:2PC-protocol}

There is today considerable research interest in the 
area of ``atomic swaps'' within public (permissionless) blockchains
(e.g. see~\cite{EzhilchelvanAldweesh2018,ZakharyAgrawal2019,Herlihy-PODC2018,ZamyatinHarz2019-XCLAIM,Nolan2013}).
However,
there is scant effort today to address the standardization of the various
infrastructure building blocks 
-- messages, data formats and flows -- 
to support the interoperability across blockchains
that permit ``swaps'' to occur.

In this section we discuss a rudimentary gateway protocol to support
the unidirectional movement of a virtual asset from an origin blockchain to a destination blockchain
with the assistance of gateway nodes at the respective endpoints.
We focus on the unidirectional movement as a building block,
with the understanding that it should be usable to achieve bi-directional conditional swaps
while satisfying the general atomicity requirements (Section~\ref{subsec:DesirableAtomicityProperties})
and security requirements (Section~\ref{subsec:DesirableSecurityProperties}).
As a common building block,
the gateway protocol must be operable in the case that one or both blockchains
are private (permissioned) systems.
For the commitment establishment between the two gateways,
we use the classic 2-Phase Commit paradigm (2PC or 3PC)~\cite{Gray1981,TraigerGray1979,OzsuValduriez2020-ddb-book},
which is a well-understood and widely deployed paradigm (e.g. in distributed database systems).

At a high level,
the gateway protocol is used to move a digital representation of an asset
from one blockchain system to another,
with the goal of maintaining consistency across the ledgers of the respective blockchains.
A gateway node must therefore interact in two directions:
(a) facing interiorly the gateway must perform the relevant tasks
in ensuring the consistency of state in its ledger,
and
(b) facing exteriorly the gateway must interact with a peer remote gateway
to obtain agreement regarding the movement of the virtual asset.

\begin{figure}[t]
\centering
\includegraphics[width=0.9\textwidth, trim={0.0cm 0.0cm 0.0cm 0.0cm}, clip]{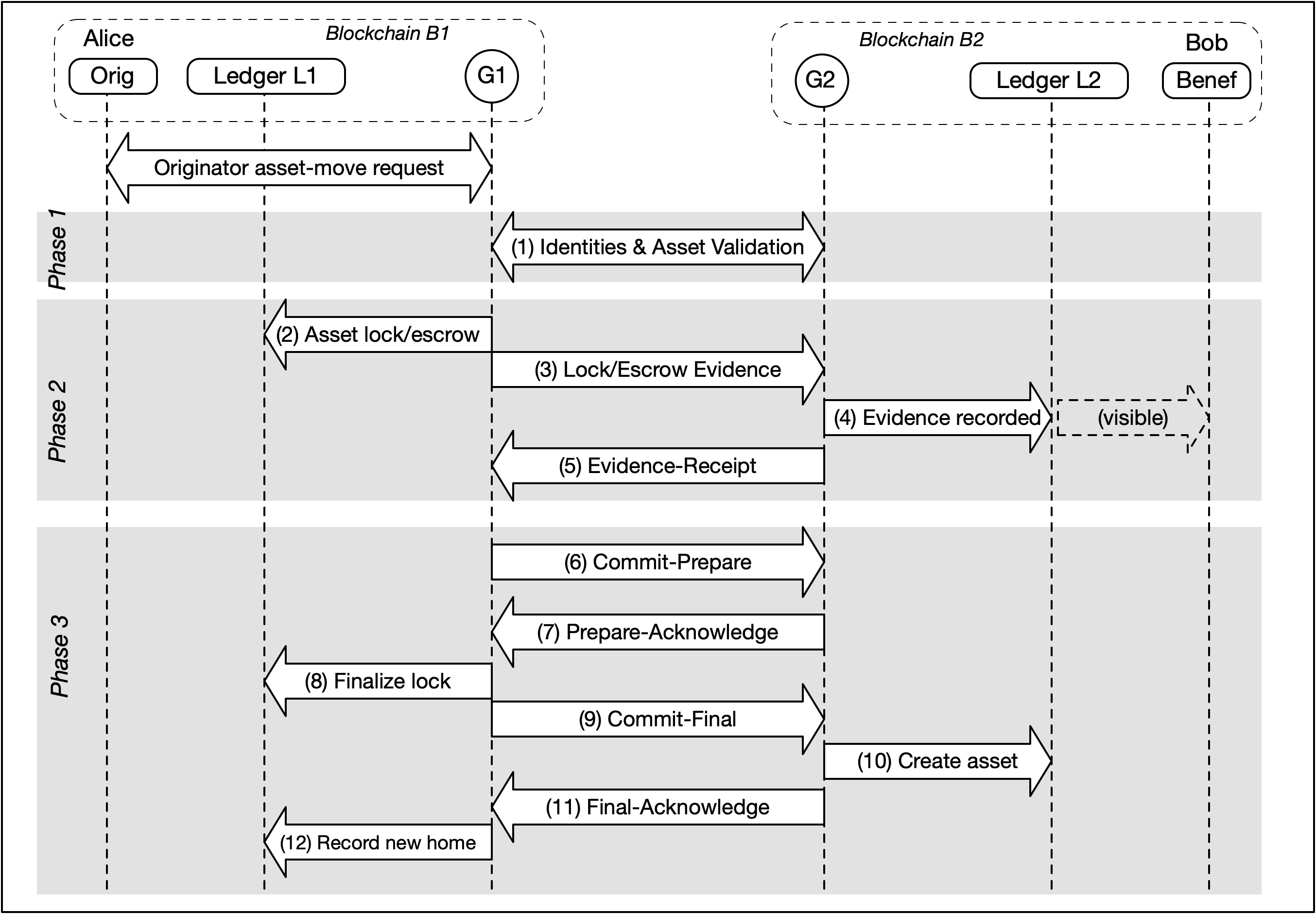}
\caption{Overview of the phases of the gateway asset movement protocol (after~\cite{HardjonoLipton2020-BookChapter,IETF-draft-hardjono-gateways})}
\label{fig:gateway-phases}
\end{figure}

In the following,
we divide the gateway protocol into three (3) general phases~\cite{HardjonoLipton2020-BookChapter,IETF-draft-hardjono-gateways}
and discuss the tasks that need to occur with in each phase.
The first phase is triggered by the originator (Alice)
in blockchain B1 seeking to move assets to the beneficiary (Bob) in B2 
(see Figure~\ref{fig:gateway-phases}).
Each gateway is assumed to keep a {\em persistent log} of all the messages
it receives or transmits to its peer gateway.
This is to permit the gateway to recover from crashes,
and to use the information in the log
to re-commence the gateway protocol.

\subsection{Phase 1: Identity Verification, Asset Validation and Parameters Exchange}
\label{subsec:Phase1}

The primary purpose of the first phase is to verify the various
information relating to the asset to be moved, the identities
of the originator and beneficiary, the identity and legal status of
the entities (VASPs) who own and operate the gateways, 
the identities of the gateways
and the exchange of session parameters needed
to establish the secure channel between G1 and G2.

This phase starts with the assumption that the gateway G1
has been selected (elected) in blockchain B1
and that gateway G2 in blockchain B2 has been identified.
The following are a summary of the various checks that need to be performed in this phase:
\begin{itemize}[topsep=8pt,itemsep=1pt, partopsep=4pt, parsep=4pt]

\item	{\em Secure channel establishment between G1 and G2}: This includes the
      mutual verification of the gateway device identities and the
      exchange of the relevant parameters for secure channel
      establishment.  In cases where device 
      attestation~\cite{HardjonoSmith2019c,HardjonoSmith2020-NodeAttest-arxiv,TCG-Attestations-Arch2020-Nov} is
      required, the mutual attestation protocol must occur between G1
      and G2 prior to proceeding to the next phase.

\item	{\em Validation of the gateway ownership}: There must be a means for
      gateway G1 and G2 to verify their respective ownerships (i.e.
      VASP1 owning G1 and VASP2 owning G2 respectively).  Examples of
      ownership verification mechanism include the chaining of the
      gateway-device X.509 certificate up to the VASP entity
      certificate, directories of gateways and VASPs, and other approaches.

\item	{\em Validation of owner legal status}: 
		There must be a means for
		gateway G1 and G2 to verify the legal status of the VASP who owns the gateway.
		In some jurisdictions limitations may
		be placed for regulated VASPs to transact only with other
      	similarly regulated VASPs ~\cite{FINMA-Guidance-2019}.
		Examples of mechanisms used to
		validate a VASP legal status include LEI numbers~\cite{GLEIF2018},
		VASP directories~\cite{Sinclair2020-coindesk}, 
		Extended Validation (EV) X.509 certificates 
		for VASPs~\cite{Hardjono2020-DETIPS-short,TRISA2020-v07}, and others.

\item	{\em Identification and validation of correct profile of asset}: 
		There must be a means for gateway G1 and G2 
		to identity and reference the definition of the virtual asset to be moved.
		We use the term {\em asset profile}~\cite{IETF-draft-sardon-assetprofile}
		to refer to this definition.
		Some jurisdictions may limit the movement of certain types of virtual assets,
		and thus it is imperative that both G1 and G2 be referring to the same asset definition.

\item	{\em Exchange of Travel Rule information}: In jurisdictions where the
      Travel Rule policies regarding originator and beneficiary
      information is enforced, the gateways G1 and G2 must exchange this
      information~\cite{FATF-Recommendation15-2018}.

\item	{\em Negotiation of asset movement protocol parameters}: Gateway G1
		and G2 must agree on the parameters to be employed within the
		asset movement protocol.  Examples include endpoints definitions
		for resources, type of commitment flows (e.g. 2PC or 3PC), 
		lock-time durations in each ledger, and others~\cite{IETF-draft-hargreaves-ODAP}.
		An important parameter is the average expected confirmation times
		of transactions within each respective blockchain network
		(e.g. the times $t_1$ in ledger L1, $t_2$ in ledger L2,
		and $t_3$ for the gateway protocol, as illustrated in Figure~\ref{fig:ThreeSegments}).
\end{itemize}

\subsection{Phase 2: Evidence of asset locking or escrow}
\label{subsec:Phase2}

In this phase, gateway G1 must provide gateway G2 with sufficient 
{\em evidence} that the asset on blockchain B1 is in a locked state 
(or escrowed) under the control of G1 on ledger L1, 
and safe from double-spend on the part of its current owner (the originator).
The purpose of this evidence is for dispute resolution between G1 and G2 
(i.e. entities who own and operate G1 and G2 respectively) 
in the case that double-spend is later detected.
Additionally,
this phase is used is prepare both gateways for the third phase where 
the actual commitment occurs.

There are several general steps which occur in this phase (see Figure~\ref{fig:gateway-phases}):
\begin{itemize}[topsep=8pt,itemsep=1pt, partopsep=4pt, parsep=4pt]

\item	{\em Asset locking in the origin blockchain}: The asset in question
must be locked (or escrowed)
to the gateway G1 and be under the exclusive control of the G1 within blockchain B1.
This is shown as Step~2 in Figure~\ref{fig:gateway-phases}.

The mechanism to lock the asset is dependent on the specific architecture of the blockchain.
For example,
a lock can be created using a passive transaction
where the gateway G1 issues a transaction on the ledger
that signals to the other nodes (e.g. mining or forging nodes)
to delay other pending transactions on the same asset
until the duration of the lock in finished.
We discuss this further in Section~\ref{subsec:ExtinguishmentRegeneration}.
Other examples of mechanisms include a direct escrow from the originator (Alice)
to the gateway G1 (i.e. to the public-key of the G1 on blockchain B1).

\item	{\em Delivery of signed lock-evidence}: The gateway G1 -- which has
the authoritative lock on the asset on in blockchain B1 --
must deliver a signed evidence to gateway G2 that the asset has been locked to G1.
The evidence must be signed using the gateway-identity key pair
which is known to the destination gateway G2 and possibly to other entities globally.
This is shown as Step~3 in Figure~\ref{fig:gateway-phases}.

The form of the evidence depends on the lock mechanism used in the origin blockchain
but should include:
(i) the transaction/block number on ledger L1 where the lock (escrow) has been confirmed;
(ii) the hash of the header of the block (assuming a Merkle Tree hash is included in the header);
(iii) the hash of the public-keys of the originator and beneficiary;
(iv) the hash of the blockchain public-key of the gateway G1;
(v) a timestamp;
and others.

The gateway G2 at the destination blockchain has the option
to passively record the received evidence on its local ledger L2
(shown as Step~4 in Figure~\ref{fig:gateway-phases}).

\item	{\em Acknowledgement of received evidence}: The gateway G2 must then
return a signed message acknowledging 
the receipt of the evidence in the previous step (Step~5 in Figure~\ref{fig:gateway-phases}).
The purpose of this acknowledgement is to provide
exculpatory evidence (cover) for gateway G1 in the case that G2 is dishonest.
As such, the message carry a hash of the previous lock-evidence message (in Step~3) 
and it must be signed
using the gateway-identity key pair of G2.

\end{itemize}

\subsection{Phase 3: Final commitment of asset movement}
\label{subsec:Phase3}

The goal of the third phase is to establish transactional commitment 
between gateway G1 and G2 that
they will each perform the relevant tasks necessary to complete the unidirectional
movement of the asset from blockchain B1 to blockchain B2.
The term {\em commitment} here is used in the sense 
of the classic 2PC model~\cite{Gray1981,TraigerGray1979}
where both sides perform actions that are durable -- following from
the atomicity properties (Section~\ref{subsec:DesirableAtomicityProperties}).
In the case of append-only ledgers,
the results of the 2PC (3PC) commitment would be recorded
by G1 and G2 in their respective local ledgers L1 and L2.

There are several steps which occur in this phase (see Figure~\ref{fig:gateway-phases}):
\begin{itemize}[topsep=8pt, itemsep=1pt, partopsep=4pt, parsep=4pt]

\item	{\em Commit preparation and acknowledgement}:
Following the 2PC (3PC) model,
gateway G1 acts as the {\em coordinator} in the 2PC commitment context.
This means that G1 has the task to ensure that the other gateway(s), namely G2 in this case,
is ready on its part to perform the commit on its ledger L2.
This is represented as Step~6 and Step~7 in Figure~\ref{fig:gateway-phases}.

\item	{\em Finalization of lock or escrow in ledger L1}:
Since the asset movement originated from blockchain B1
-- with gateway G1 being the commitment coordinator --
the finalization of the lock (escrow) must occur first on ledger L1.
This means that in Step~8 
gateway G1 must record on ledger L1
that the asset has been moved to the address (public key)
of the beneficiary located on ledger L2 in blockchain B2
with the assistance of gateway G2.
We discuss further
passive transactions and escrows  in Section~\ref{sec:Dicussion}.

\item	{\em Delivery of evidence of final lock/escrow}:
Once the lock (escrow) finalization has been recorded (confirmed) on ledger L1,
the gateway G1 must deliver evidence of this confirmed lock/escrow on L1
to the gateway G2.\\
Because of the {\em opaque ledgers assumption} (see Section~\ref{subsec:DesignPrinciples}),
gateway G2 has no visibility into ledger L1 and therefore must solely trust
the signed assertion by gateway G1 that the
lock (escrow) finalization has been confirmed on L1.
The signed assertion delivered in Step~9 from G1 to G2
must include the entire information (data)
confirmed on ledger L1 from the previous Step~8 -- even though gateway G2
has no way to validate the truthfulness of the information.
The purpose of this is to provide exculpatory legal evidence for gateway G2
(and its owner/operator VASP2) in the case
that gateway G1 cheats and that a duplication (double-spending) of the asset has occurred.

\item	{\em Finalization of asset regeneration in ledger L2}:
Based on the signed assertion from G1 delivered in Step~9,
the gateway G2 performs a similar finalization of the 
asset regeneration on ledger L2 in blockchain B2.
The information recorded on L2 must include
the signed information/data (or hashes of) received by G2,
as a means to protect G2 (i.e. as exculpatory proof) against a dishonest G1 or a dishonest beneficiary.
This is shown as Step~10 in Figure~\ref{fig:gateway-phases}.

Depending the data structure (i.e. block structure) of the ledger L2,
gateway G2 may have to map the data from the signed assertion
to the format used on L2.
If the block size on L2 cannot accommodate the full assertion data from G1 (from the previous step),
then a hash of the assertion should be used instead -- with the full signed assertion
being stored off-chain in protected storage.

\item	{\em Report asset regeneration commitment (acknowledge-final)}:
Once the asset regeneration transaction is confirmed on ledger L2,
the gateway G2 may optionally report the new block/transaction details (on L2)
to gateway G1.
This is shown as Step~11 in Figure~\ref{fig:gateway-phases}.

This step is not strictly part of the 2PC model,
but in the context of asset movements across private (permissioned) blockchains
it provides a means for future tracing of the asset movements
from the legal perspective (e.g. legal audit).

\item	{\em Local recording of new asset location}:
Assuming gateway G2 provides a report to G1 of the new location of the asset
(i.e. new block/transaction details on L2),
the gateway G1 may record this data on ledger L1.

\end{itemize}

\section{Discussion}
\label{sec:Dicussion}

\begin{figure}[t]
\centering
\includegraphics[width=1.0\textwidth, trim={0.0cm 0.0cm 0.0cm 0.0cm}, clip]{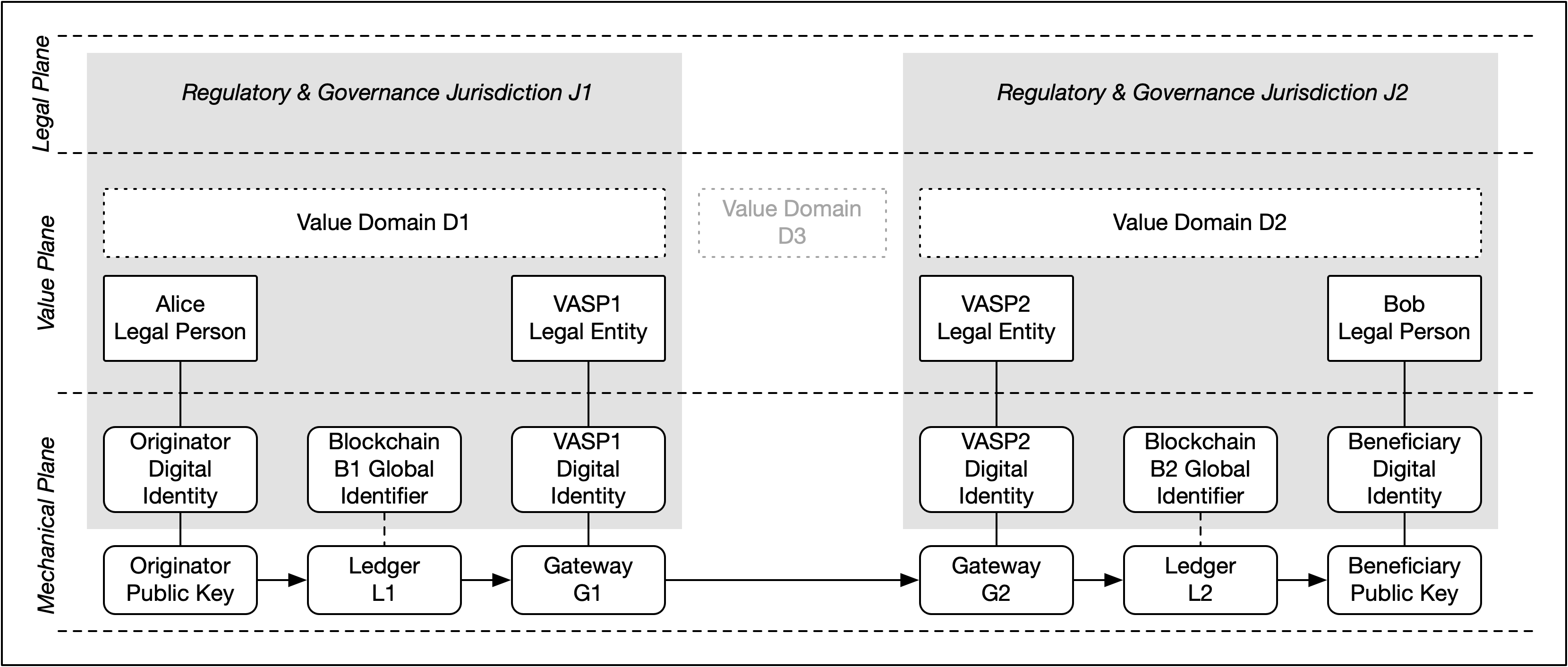}
\caption{Asset Movement at the Mechanical, Value and Legal Jurisdictions Planes}
\label{fig:threeplanes}
\end{figure}

The movement of virtual assets between blockchain networks
is a complex problem because it straddles (cuts across)
three (3)  ``planes'' (or  domains)
that together provide an interpretation of economic value of the virtual assets
according to the prevailing socio-economic conventions
and legal/governance rules (regulations) in a given area of the world.
We refer to these three planes as:
(i) the mechanical (technical protocols) plane,
(ii) the value plane (i.e. economic or financial values)
(iii) the legal and governance jurisdictions plane
(see Figure~\ref{fig:threeplanes}).

In the following we discuss various aspects and challenges
related to the gateway architecture and gateway-protocol
described above.

\subsection{Extinguishment and Regeneration of Assets: Locks \& Escrows}
\label{subsec:ExtinguishmentRegeneration}

As mentioned previously in Section~\ref{subsec:ThreeParts},
at the {\em mechanical plane} (protocol layer)
the proposed gateway architecture implements the notion of asset extinguishment and regeneration.
Thus, at the protocol layer we define the following:
\begin{quote}
	A unidirectional cross-blockchain 
	movement of virtual assets consists of the 
	{\em extinguishment} of the asset representation
	in the ledger of the origin blockchain
	and for the {\em regeneration} of the asset representation 
	within the ledger of the destination blockchain,
	conducted as an {\em atomic} operation.
\end{quote}
There several technical means to realize the notion of ``extinguishment'' and ``regeneration''
of virtual assets on distributed ledgers,
with the meditation of gateway nodes.
These two operations must be performed atomically,
leading to consistency of data across both ledgers.

The atomicity-related requirements (Section~\ref{subsec:DesirableAtomicityProperties}) 
necessitate the use of temporary ``locks'' on
the asset in question on the origin blockchain
to temporarily disable it and prevent double-spending.
This introduces the question of how ``locks'' can be employed on ledgers
which by definition are append-only data structures~\cite{HaberStornetta1991,BayerHaber1993}.
The ledger recording of a temporary lock
must be communicated by gateway G1 in the origin blockchain
to gateway G2 at the destination blockchain
as a form of evidence that the atomic movement of the asset has commenced.

Two possible mechanisms for temporary locks or disablement of asset
are as follows:
\begin{itemize}[topsep=8pt, itemsep=1pt, partopsep=4pt, parsep=4pt]

\item	{\em Conditional escrow transaction}: An escrow transaction on an asset on a blockchain
alters the ownership of the asset
from its current owner to a trusted third party based on some condition.
An escrow can be {\em timed} in that a duration of time $t_{esc}$
is allocated within which the condition must be fulfilled.
In the context of cross-blockchain asset movement,
the condition is related to the atomic commitment mediated by gateway G1 and G2
through their execution of the gateway-protocol.
If the gateway protocol finishes to completion within the set duration of time $t_{esc}$,
then the escrow is finalized
and the asset reverts ownership.
If the gateway protocol fails to conclude within time $t_{esc}$,
then the asset reverts back to the originator.

The precise method to implement escrows is dependent on the protocols employed
by the blockchain (e.g. combinations of hash-locks, time-locks, multisig, smart contracts, etc)
and the asset representation on the ledger (e.g. fungible, non-fungible).
However,
regardless of the precise mechanism, 
the escrow {\em must be confirmed on the ledger} and therefore be visible to the all the entities
having access to the ledger.

\item	{\em Passive blocking \& disablement transaction}: A passive blocking transaction 
(directed at an asset) is a type of transaction that refers to the asset on the ledger
for the purpose of temporarily blocking or deferring other actions on the same asset
by other nodes/trasactions.
The passive blocking transaction contains a time duration $t_{block}$
that counts once the transaction has been confirmed on the ledger.
It does not change the ownership of the asset 
(i.e. the public-key most recently associated with the asset).

A passive blocking transaction can either time-out (i.e. blocking no longer valid)
or it can be made permanent by a matching {\em passive disablement transaction}
that makes the effect of the blocking transaction to be permanent.
Once the disablement transaction is confirmed on the ledger,
the asset in question is no longer considered to be active
on the blockchain.

\end{itemize}
Both types of mechanisms provide a means to temporarily ``lock'' an asset
with the subsequent option to make the lock effect permanent on the ledger 
(i.e. extinguish the asset by making it henceforth inaccessible).
Furthermore,
in both approaches there is evidence of these actions recorded (appended) permanently on the ledger.
As an illustration,
the Phase-2 of the gateway protocol (Section~\ref{subsec:Phase2} and Figure~\ref{fig:gateway-phases})
employs a pair of ``locks'' on ledger L1.
In Step~2 the gateway G1 issues a temporary asset disablement,
which is subsequently finalized 
in Step~8 of Figure~\ref{fig:gateway-phases}.
If Step~2 is never closed by a matching Step~8 (e.g. due to gateway crashes),
then the lock effect will time-out and will be ignored by other nodes.


Symmetrically to the disablement of an asset at the origin blockchain,
the same asset must be introduced into the destination blockchain
in two basic steps, namely the {\em notification} step and the {\em finalization} step performed
by the gateway G2.
The notification step ``announces the impending arrival'' of an asset to be assigned
to the beneficiary's address (public-key) on the ledger L2 of the destination blockchain.
This notification transaction is passive in that it acts
as a log entry to be recored on the ledger L2.
It does not assign ownership of the asset to the beneficiary.
Ideally,
the passive {\em notification transaction} issued by G2 
should include the the lock/disablement information delivered by gateway G1 to G2,
the relevant timestamps,
and the address (public key) of the beneficiary.

In order for the asset to be fully regenerated in blockchain B2,
a {\em finalization transaction} must be issued by G2 that formally
records on ledger L2 the ownership assignment of the asset to the beneficiary's public-key.
The finalization transaction must include relevant information,
including a hash of the previous notification transaction
confirmed on ledger L2,
as a means to refer to the specific asset movement in question. 
Continuing the above illustration,
the passive notification transaction is shown as Step~4 within the gateway protocol 
(Section~\ref{subsec:Phase2} and Figure~\ref{fig:gateway-phases}).
The finalization transaction is shown as Step~10 within the gateway protocol
(Section~\ref{subsec:Phase3} and Figure~\ref{fig:gateway-phases}).
If Step~4 is never closed by a matching Step~10 (e.g. because gateway crashes),
then the announcement transaction will time-out and be thereafter ignored.

\subsection{Gateway Identities, Signing Keys and Certificates}
\label{subsec:GatewayIdentitiesKeys}

\begin{figure}[t]
\centering
\includegraphics[width=0.9\textwidth, trim={0.0cm 0.0cm 0.0cm 0.0cm}, clip]{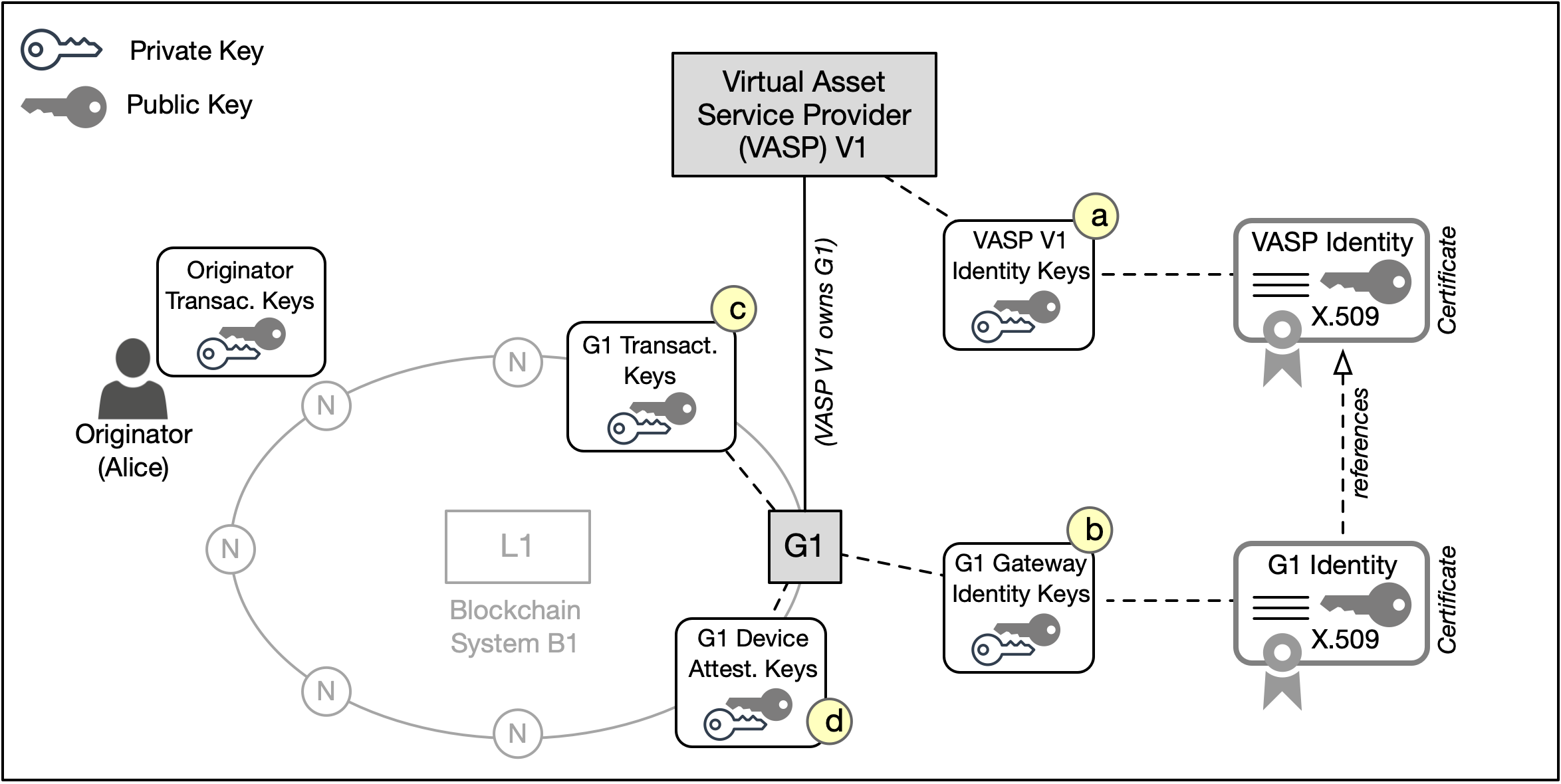}
\caption{Gateway identity and VASP identity key-pairs \& certificates}
\label{fig:deviceidentities}
\end{figure}

In jointly executing the gateway protocol (Section~\ref{subsec:2PC-protocol}),
the gateways G1 and G2 must perform a number of digital signatures
on the messages in the protocol.
These digital signatures signify the promise or commitment on the part of the owners
of the gateways to perform the asset movement following the prescribed protocol.
The signatures becomes crucial in the case of disputes occurring
between the owners of the gateways (and possibly involving the originator and beneficiary).
As such,
mechanisms are needed to cryptographically bind the public-key of a gateway to 
the identity of its owner (a VASP) in such a way that 
the digital signatures~\cite{esign2000misc} performed by a gateway
becomes legally binding to its owner.
Figure~\ref{fig:threeplanes} provides a high level illustration of the entity identities,
which we have expanded upon in Figure~\ref{fig:deviceidentities}.

We define at least four (4) types of public-private key pairs
associated with a gateway and its owner (Figure~\ref{fig:deviceidentities}):
\begin{itemize}[topsep=8pt, itemsep=1pt, partopsep=4pt, parsep=4pt]

\item	{\em VASP identity key-pair}: This public-private key pair is used by 
the VASP in interacting (off-chain) with other entities
in the capacity of the VASP as a legal entity.
This is illustrated in Figure~\ref{fig:deviceidentities}(a).
The VASP should employ a different key-pair when
signing transactions on a blockchain system.
The VASP identity public-key may be represented
inside an Extended Validation (EV) {X.509} certificate
(see~\cite{Hardjono2020-DETIPS-short,TRISA2020-v07} for a discussion of EV-Certificates for VASPs).

\item	{\em Gateway identity key-pair}: This public-private key pair is used
by a gateway when interacting with peer remote gateways (i.e. facing outbound).
Among others, it is used to set-up the secure channel with the peer gateway
and to sign the various protocol messages exchanged between the two gateways.
This is shown in Figure~\ref{fig:deviceidentities}(b).

Ideally, the gateway identity public-key should be represented within
the standard {X.509} certificate,
with additional fields containing either the hash of the VASP identity public-key
(or hash of VASP identity certificate), or the 
serial-number VASP identity certificate.
The gateway {X.509} certificate should be issued under the VASP
{X.509} identify certificate.
This certificate linking or chaining signifies the ownership of the gateway by the VASP.

\item	{\em Gateway transaction signing key-pair}: This public-private key pair is used by
the gateway to sign transactions on its interior ledger in the gateway's capacity
as a node in the blockchain network.
Depending on the requirements of the blockchain,
the public-key may be represented within an {X.509} certificate 
that is bound to the VASP identity certificate.
This is shown in Figure~\ref{fig:deviceidentities}(c).

\item	{\em Gateway device attestations key-pair}:
This public-private key pair is a device-level key pair
used by the gateway to sign attestations evidence~\cite{TCG-Attestations-Arch2020}
pertaining to the hardware/software stack~\cite{NIST-800-193} underlying the gateway implementation
(see~\cite{HardjonoSmith2020-NodeAttest-arxiv} for a discussion of node attestations
within blockchain networks).
Ideally the key pair should be bound to the hardware
of the gateway in such a way that it is difficult (un-economical)
for attackers to physically remove, export or replace the private key.
Solutions for hardware-bound keys have been developed
by several industry organizations~\cite{TPM1.2specification,HardjonoKazmierczak2008b,TPM2.0specification,IEEE-DeviceID}.
This is shown in Figure~\ref{fig:deviceidentities}(d).

\end{itemize}

\subsection{A Certificate Hierarchy for VASPs and Gateways}
\label{subsec:CertificateHierarchy}

When a gateway opens a connection with 
a peer remote gateway in another blockchain system,
one immediate requirement of the respective gateways 
is the verification of the legal ownership of the gateway device
(i.e. which VASP owns the gateway) and 
the validation of the legal status of the VASP (see Section~\ref{subsec:Phase1}).
We use the generic term ``VASP'' for the owner and operator of gateways
following the FATF definition~\cite{Hardjono2019b,FATF-Recommendation15-2018}.

\begin{figure}[t]
\centering
\includegraphics[width=0.9\textwidth, trim={0.0cm 0.0cm 0.0cm 0.0cm}, clip]{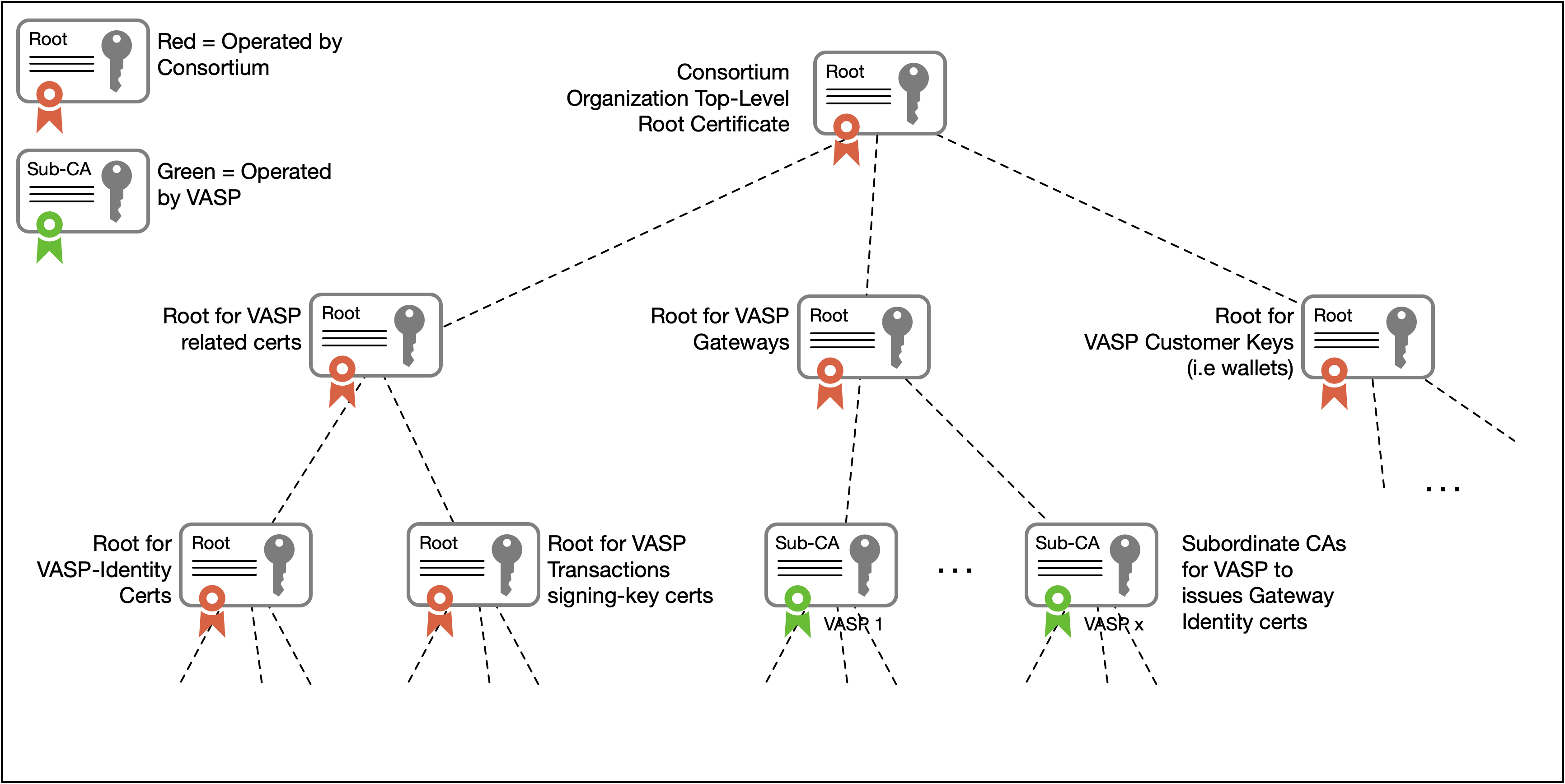}
\caption{Example of a certificate hierarchy for VASP identities and their gateways}
\label{fig:CertHierarchies}
\end{figure}

One possible solution is for gateways to employ {X.509} digital certificates~\cite{rfc5280,ISO9594-pubkey}
and for the certificates to be exchanged as part of the TLS handshake~\cite{RFC5246-formatted,RFC8446-formatted}.
This approach of exchanging {X.509} certificates is a common practice 
in the Internet online commerce industry since the mid-2000's.
The CA/Browser (CAB) Forum has additionally defined
{\em Extended Validation} (EV) fields,
which contains the business-related information of the subject (i.e. business entity)~\cite{CAB-Forum2020}.
A similar EV-certificates approach is being adopted currently for VASPs,
where the VASP business information
(e.g. business registration, legal address, LEI number, etc.) 
is added as the EV fields in their {X.509} certificates~\cite{Hardjono2020-DETIPS-short,TRISA2020-v07}.

In the case of gateway identity public keys,
we believe a {\em certificate hierarchy} is required that can associate or bind a gateway 
{X.509} identity certificate to the owner's {X.509} EV-identity certificate,
and which permits any entity to quickly perform ``chain validation'' (up the certificate chain) of these certificates.
Such a device/operator combined certificate hierarchy 
is not new, and has been deployed by the Cable Modem industry for over two decades now
under the industry consortium called Cable Laboratories (Cable Labs)~\cite{CableLabs-CPS-2019,CableLabs-CertIssuance-2010}.
This solution was developed to permit DOCSIS1.2~\cite{DOCSIS3.1-Security} (or higher)
cable modem devices to perform mutual authentication with its head-end device
(cable modem termination service or CMTS) that is owned by the regional
cable provider/operator company.
A similar approach for WiFi and Wireless-LAN devices is described in~\cite{HardjonoDondeti2005,RFC3770-Formatted}.

Figure~\ref{fig:CertHierarchies} provides an illustration for a potential certificate hierarchy,
that is rooted in an industry consortium.
One purpose of the certificate hierarchy is to permit
recipients of any VASP-related certificate (issued under the hierarchy) 
to be able to quickly validate the certificate,
with the assurance that the VASP legal status has been verified by the consortium~\cite{TRISA2020-v07}.
Another purpose is to permit VASP the freedom to issue some relevant certificates,
such as the gateway identity certificates (for their gateways)
and customer signing key certificates (for their customers whose keys are hosted by the VASP).

A deeper discussion of VASP certificates and gateway certificates
is outside the scope of the current work,
and will be treated in future work.

\subsection{Gateway Crash Recovery}
\label{subsec:CrashRecovery}

An important issue with the use of gateways is that of the potential unavailability
of a gateway, due to crashes or other availability issues.
In order to anticipate gateway crashes,
a crash recovery scheme or strategy should be part of the design of gateways.
This design should also be informed by the blockchain architecture
for which the gateway is designed (i.e. inward facing).
The overall goal of a crash recovery scheme
is to maintain the desirable atomicity properties (Section~\ref{subsec:DesirableAtomicityProperties})
and security properties (Section~\ref{subsec:DesirableSecurityProperties}).
As mentioned previously,
a gateway must retain a persistent log of all the messages that it received from (and sent to)
its peer gateway in the execution of the asset movement protocol (e.g. ODAP protocol~\cite{IETF-draft-hargreaves-ODAP}).

There are several aspects related to crash recovery
that needs to be taken into consideration:
\begin{itemize}[topsep=8pt, itemsep=1pt, partopsep=4pt, parsep=4pt]

\item	{\em Recovery gateway selection}: 
There needs to be a mechanism to determine
the {\em recovery gateway} that 
will re-engage the remote peer gateway to complete 
the gateway protocol (asset movement) session.
There are several approaches or strategies that can be adopted.
One approach could be {\em self-healing},
meaning that the same gateway device recovers (e.g. reboots)
and picks-up the session from the last possible check-point in the log.
Another approach could be the use of {\em primary backup} gateway
that detect the crash of the primary gateway and takes-over
the re-starting of the gateway protocol~\cite{IETF-draft-belchior-crashrecovery}.
An alternative strategy could be election
of the recovery gateway,
such as through the use of the consensus algorithm itself.

\item	{\em Reliable storage of event logs and access to logs}:
There needs to be storage facility to store logs
of the messages received from (sent to) a peer gateway,
such that the entire sequence can be reconstructed up to the point 
where the crash occurred.
There are several possible approaches with regards 
to the storage of logs by a gateway in a blockchain network.
This includes (i) placing the log data on the main ledger (i.e. transactions ledger),
(ii) storing the log data off-chain, with each entry hashed to the main ledger (for data integrity purposes),
(iii) establishing a special ``metadata'' ledger within the blockchain network
solely for the purpose of recording the log data,
and other approaches.

A key requirement is that the log entries be accessible (i.e. readable)
to the recovery-gateway.
A standardized API to write to (read from) the log
should be defined,
where the implementation of the log mechanism (behind the API)
can be a local construct~\cite{Belchior2021-Hermes-long}.

\item	{\em Secure channel re-establishment}:
A recovery gateway must establish a secure channel with
its peer remote gateway in order to complete the the gateway protocol (asset movement) session.
If the recovery gateway is the same device
as the original primary gateway (i.e. self-healing node),
then depending on the secure channel protocol (e.g. TLS, IPsec, etc.) and its configuration
the recovery gateway may be able to resume the secure channel (i.e. TLS session resumption).
However,
if the recovery gateway is different device from the original gateway,
then a new secure channel must be created with the remote peer gateway.

\end{itemize}

\subsection{Gateway Discoverability}
\label{subsec:GatewayDiscovery}

The ability for a gateway G1 (fronting blockchain B1) to discover its 
peer gateway G2 (fronting blockchain B2) 
is closely related to the current fundamental challenge
of VASP discovery in the virtual assets industry~\cite{TRISA2020-v07}.

More broadly,
the issue of originator and beneficiary discoverability
is relevant from the Travel Rule compliance perspective~\cite{FATF-12MonthReview-2020}
because an Originator-VASP must obtain and validate 
the personal data of the beneficiary and the legal status of the Beneficiary-VASP
prior to transmitting assets to the beneficiary.
Compliance to the funds Travel Rule has been the norm
for the banking sector and correspondent banking since 
the Bank Secrecy Act of 1996 (BSA - 31~USC~5311-5330).
Not only do VASPs -- such as crypto-exchanges -- need to comply to the Travel Rule,
they also have the additional burden related to proving control over private-keys~\cite{HardjonoLipton2020-FinTech},
which are the means to control virtual assets on a blockchain network.
Thus, the discoverability of VASPs and their legal status
permits the question of the legal ownership of gateway nodes 
(and control over gateways) to be answered
more succinctly
at all three planes of the asset movement paradigm summarized previously in Figure~\ref{fig:threeplanes}.

The following is a short list of some of the challenges and requirements
related to the problem of the discoverability of gateways
at a global scale.
For gateway G1 to discover the correct gateway G2 (and its protocol endpoints),
we assume that the decision to perform the asset movement 
(from the originator in B1 to the beneficiary in B2)
has been taken at the value plane.
\begin{itemize}[topsep=8pt,itemsep=1pt, partopsep=4pt, parsep=4pt]

\item	{\em Persistent global identifier for gateway devices}:
In order for one gateway to discover another,
each gateway device must have (be assigned) a persistent identifier
that is globally unique and which survives crashes and re-starts.
The gateway identifier must be independent from 
the network addressing mechanism (i.e. IP address) of the gateway device,
because network-layer addresses may change.
It must also be independent from the identity public/private key-pairs
because these keys maybe replaced over time (e.g. aging keys being archived).
The gateway identifier could be initially cryptographically derived
from its keys (e.g. see~\cite{rfc3972}),
but its usage must not be dependent on the existence of the keys.
The persistency of the identifier is needed also
from a regulatory audit perspective,
where past logs may refer to a gateway identifier 
and the gateway owner
even though the gateway may no longer be operational (e.g. dead hardware).

There are several persistent-identifier assignment paradigms
that can be adopted,
such as the DOI~\cite{KahnWilensky1995},
KERI~\cite{Smith2019-KERI},
and
hardware-derived identifiers
(e.g. derived from EK public-key of TPM hardware~\cite{TPM1.2specification,HardjonoKazmierczak2008},
derived from hardware key ``latches''~\cite{England-CyRep-Microsoft-2017,TCG-DICE-Hardware-Requirements-2018}, etc.).

\item	{\em Binding between gateway identifier and network-layer address}:
There needs to be a mechanism to bind the persistent gateway identifier
with network-layer address of the gateway (i.e. IP address).

Example mechanism include using DNS/DNSSEC records~\cite{rfc4255,rfc4870,rfc6376},
Handle system records~\cite{rfc3650,rfc3651,rfc3652},
DID data structures~\cite{W3C-DID-2018},
{X.509} certificates,
and others.
Approaches such as the VASP Directory proposal of~\cite{TRISA2020-v07}
can be augmented to include the identifiers of gateways owned by each VASP in the ecosystem.

\item	{\em Resolver services}:
Corresponding to the binding between the gateway identifier and network-layer address,
there needs to be matching resolver services
that can ``map'' from a gateway identifier to its current IP address.
Examples of resolver services include DNS/DNSSEC~\cite{rfc4255,rfc4870,rfc6376},
Handle system~\cite{rfc3652},
and others.

\end{itemize}

\section{Looking Further Afield}
\label{sec:FurtherAfield}

In this section,
we discuss a number of topics relating to gateways,
potential variations of the gateway architecture,
and the use of external constructs -- such as a witness blockchains -- to provide
greater flexibility to the basic gateway model defined in Section~\ref{sec:GatewayArchitecture}.

\subsection{Third Party Witness Blockchains}
\label{subsec:WitnessBlockchain}

There has been several authors who have pointed out 
the benefit of a public {\em witness blockchain} that acts
purely as an append-only blockchain that can carry
any data of fixed length and which can be ``written to''
by anyone (e.g.see~\cite{ZakharyAgrawal2019,HerlihyLiskov2019,TradecoinRSOS2018}.
Indeed, this idea harks back to the purpose of the rudimentary hash-chain
model by Haber and Stornetta in 1991~\cite{HaberStornetta1991,BayerHaber1993},
which included the periodic printing in a local 
newspaper of the roots of hash-tree (e.g. hexadecimal strings).
Thus, the main purpose of such a witness blockchain is 
to act as a decentralized notarization service,
where pairs of {\tt <public-key,data>} with a timestamp
could be recorded on the append-only public ledger.

In the context of the gateway protocol jointly executed
between two gateways G1 and G2,
such a public witness blockchain can be utilized in several ways:
\begin{itemize}[topsep=8pt,itemsep=1pt, partopsep=4pt, parsep=4pt]

\item	{\em Summary log for gateway protocol messages}:
A witness blockchain could be used to store the hash of the messages
exchanged between G1 and G2 as part of the gateway protocol
(see Section~\ref{subsec:Phase2} and Section~\ref{subsec:Phase3}).
The witness blockchain becomes a notarized storage of
evidence in the case of future disputes between G1 and G2
(i.e. their owners VASP1 and VASP2 respectively).

\item	{\em Forwarding address for moved assets}:
The notion of a forwarding address was suggested in~\cite{HardjonoLipton-IEEETEMS-2019}
as a means for assist in the search for the ``current home'' of certain types of virtual assets.
This is notably relevant in the case that both blockchains B1 and B2 are private/permissioned,
and where the asset is a non-fungible asset
(e.g. ownership of a real-estate property, 
ownership certificates of commodities, etc.)~\cite{DTCC-Project-Whitney-2020,MAS-Singapore-Ubin5-2020}.

Thus, looking at Figure~\ref{fig:gateway-phases},
the last step (Step~12) 
in Phase~3 is the recording to ledger L1 of the asset's new location at blockchain B2.
Here,
gateway G1 (or G2) may additionally record a subset of the information from Step~12
onto a public witness blockchain.
The information could consists, among others, of: 
(i) the asset type-identifier,
(ii) the hash of the digital representation of the asset (e.g. non-fungible),
(iii) the hash of the beneficiary public-key,
and 
(iv) the blockchain system identifier (e.g. B2).

This permits a future look-up by an external querier for the asset ownership at B1
to be redirected to the witness blockchain,
where the querier will discover the latest location of the asset (now at B2).

\end{itemize}

\begin{figure}[t]
\centering
\includegraphics[width=1.0\textwidth, trim={0.0cm 0.0cm 0.0cm 0.0cm}, clip]{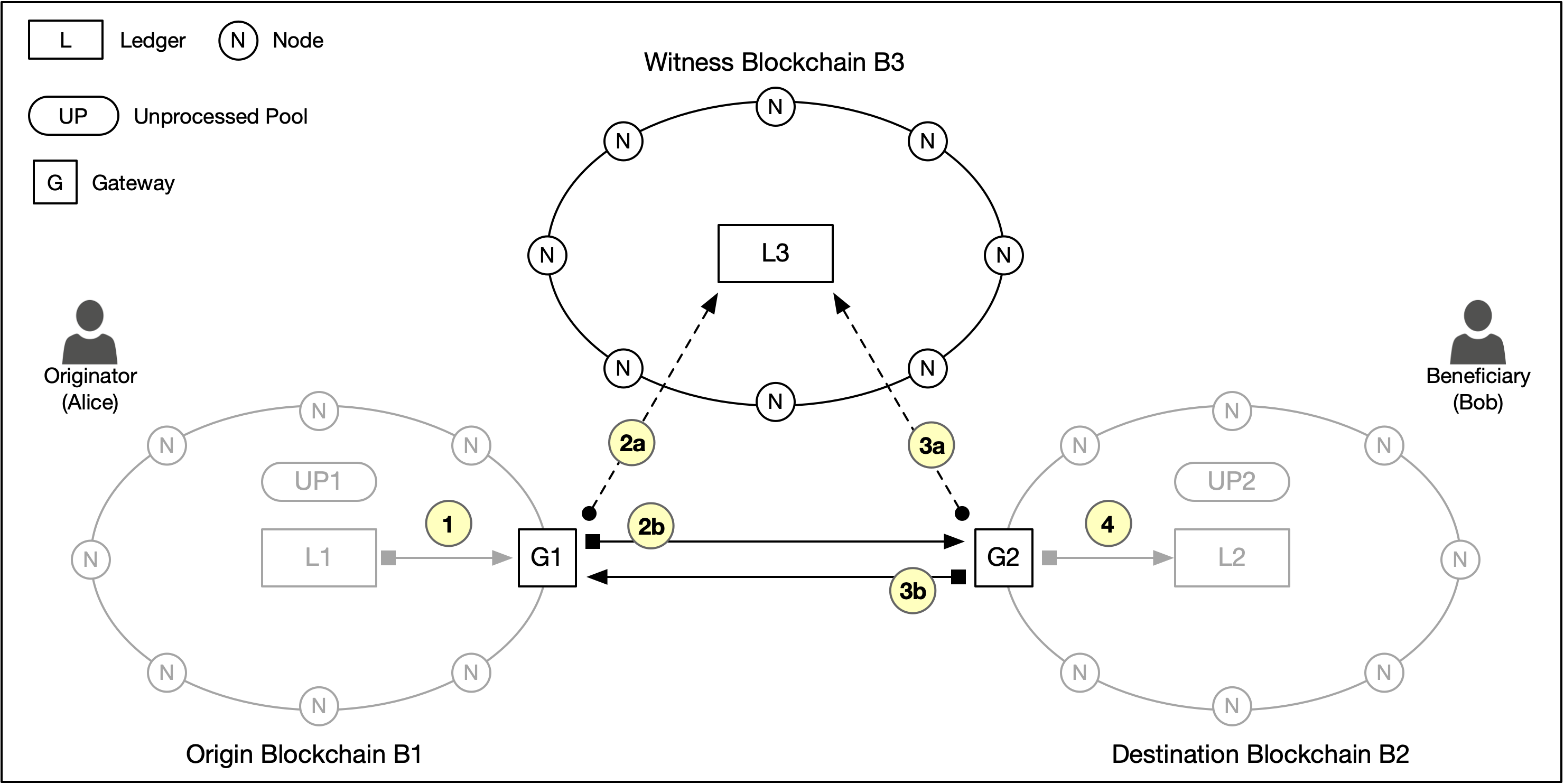}
\caption{The notion of a public witness blockchain}
\label{fig:witnessblockchain}
\end{figure}

\subsection{Bridges for Private Blockchain Networks}
\label{subsec:Bridges}

We define a {\em blockchain bridge} as a partial gateway
in a private blockchain network that is authorized
to expose (export) a number of data elements from its interior ledger
under the authorization of a user (e.g. originator)
in the context of a cross-chain asset movement initiated by the user.
Thus, a bride node provides an {\em information conduit} function only,
and not a transactional function in the sense of a full gateway.
Unlike a full gateway,
a bridge node can only read from (not write to)
its interior ledger.

In the context of a cross-blockchain asset movement,
a bridge node may be utilized by an originator (Alice) to provide evidence
to an external entity (e.g. beneficiary Bob) regarding the originator's ownership of a given asset,
without giving the beneficiary access to the private ledger.
Such evidence may be needed before the first phase of the gateway protocol (Section~\ref{subsec:Phase1})
commences.
Since the originator (e.g. user) maybe prohibited by the governance of the private blockchain (e.g. consortium)
to share ledger data with external entities,
the bridge construct is one way to implement the authorized sharing of redacted transaction data from the ledger
with a specified external entity (such as the beneficiary).
The location of the exported ledger-data should be pre-approved by the governance of the private blockchain,
and could include a public witness blockchain (Section~\ref{subsec:WitnessBlockchain}).
Several mechanism can be used to elect the bridge node,
including a consensus protocol instance.

Using the example of the preparation for a cross-blockchain asset movement
from Alice (originator) to Bob (beneficiary),
the following are some tasks of a bridge node:
\begin{itemize}[topsep=8pt,itemsep=1pt, partopsep=4pt, parsep=4pt]

\item	{\em Redaction of transaction data}:
The bridge node must remove certain fields of information
that may disclose information regarding other interior entities
unrelated to the current cross-blockchain asset movement.

For example,
if the past transaction history on the ledger L1 indicates that
the last local asset movement was from Charlie's public-key to Alice's public-key
(signifying that Alice is now the current holder of the asset),
then the bridge node must redact (remove) Charlie's public-key from the information
it will export.
This is because Charlie's public-key has no bearing
to the current the cross-blockchain asset movement
from Alice in blockchain B1 to Bob in blockchain B2.

In the case that the local ledger L1 in blockchain B1
is protected using data encryption techniques
(e.g. Zero-Knowledge Proof schemes~\cite{ZeroCash2014},
or encrypted private channels~\cite{Androulaki2018-FabricSecureChannels})
then the bridge node must first decipher the relevant block/transaction
before redacting it (assuming it has the capability and keys to decrypt these).

\item	{\em Encryption of exported data}:
In order to protect the privacy of the information being exported
(e.g. the fact of Alice in blockchain B1 being the current asset owner),
the bridge node can apply encryption to the data prior to exporting
to a public witness blockchain.

For example,
the exported data could be encrypted to the beneficiary's public-key
on the witness blockchain,
permitting only the beneficiary to view the data.

\item	{\em Retain authorization tokens}:
The owner of a bridge node (i.e. VASP)
must be protected against future disputes from a dishonest originator
(e.g. Alice repudiates the authorization to export her ledger-data).
Thus, the bridge node must also retain and archive
the authorization-token (e.g. OAuth2.0 tokens~\cite{UMACORE1.0,KantaraBSC2017})
used by the originator to authorize the bridge node.

\end{itemize}
The topic of user authorization and consent for a bridge node
to export ledger data is beyond the scope of the current work,
and will be treated in future work.

\subsection{Gateways and Delegated Hash-Locks}
\label{subsec:DelegatedHashLocks}

\begin{figure}[t]
\centering
\includegraphics[width=1.0\textwidth, trim={0.0cm 0.0cm 0.0cm 0.0cm}, clip]{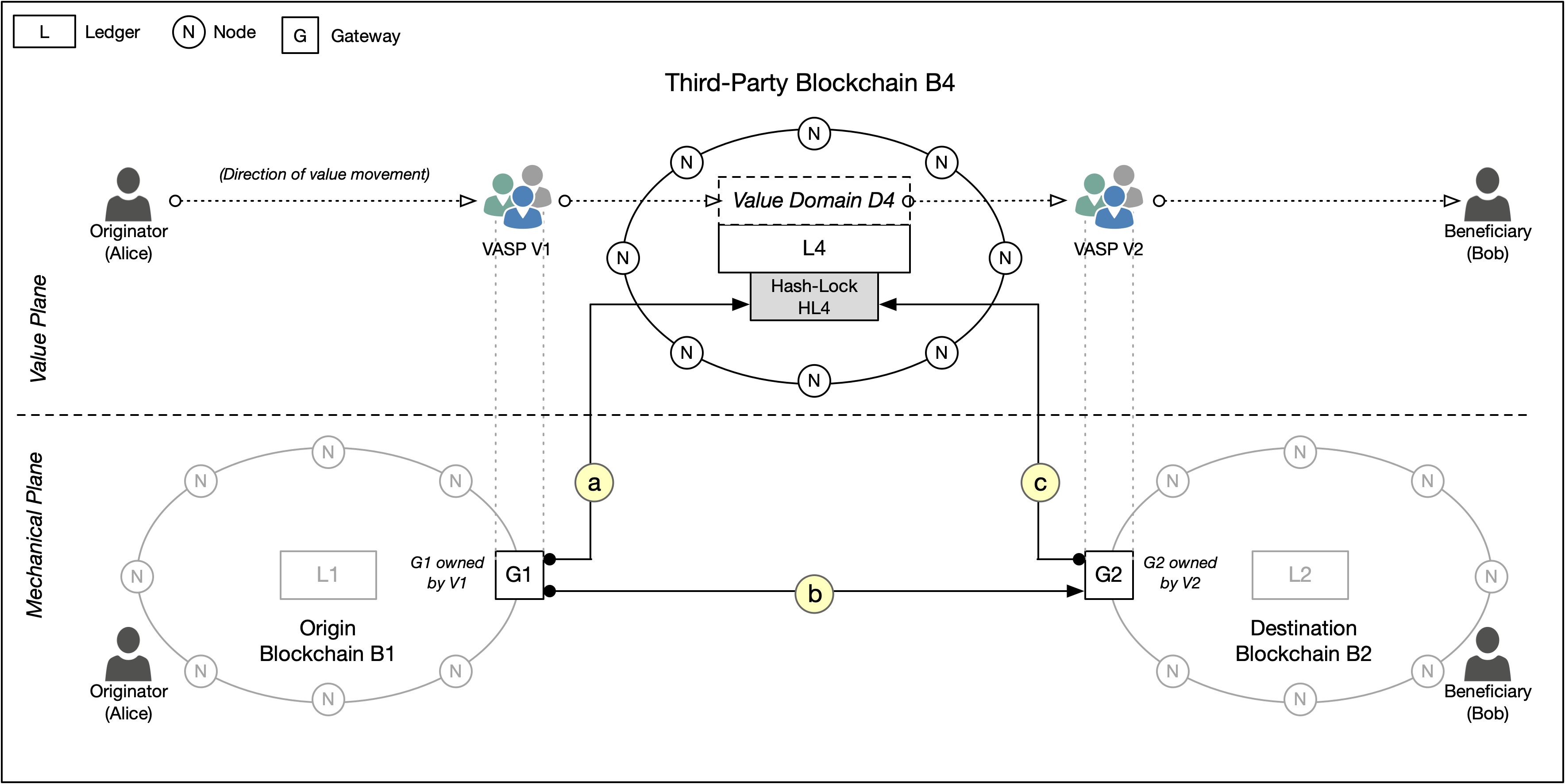}
\caption{Overview of a delegated remote hash-lock by a gateway}
\label{fig:remote-hash-lock}
\end{figure}

One potential function that can be implemented by a gateway
is accessing a {\em Hash Time Lock Contracts} (HTLC) -- commonly
referred to as ``hash-lock''~\cite{Nolan2013} smart contracts -- 
at a third-party remote blockchain 
(denoted as B4 in Figure~\ref{fig:remote-hash-lock}) to which the gateway has access.
This approach maybe useful for cases involving fungible\footnote{The Merriam-Webster dictionary 
defines {\em fungible}  as being something (such as money or a commodity) 
of such a nature that one part or quantity may be replaced by 
another equal part or quantity in paying a debt or settling an account.}
assets in which the cross-blockchain movement involves
the exchange of value to a common denomination (e.g. fiat-backed stablecoin)
occurring simultaneously at remote blockchain B4.
A hash-lock~\cite{Nolan2013} is defined as $h = H(s)$ where $H$ is a strong
cryptographic hash function and $s$ is the secret to unlock it within the duration time $t_h$.
A hash-lock can be implemented using a smart contract on a blockchain system
which enforces the timeout $t_h$.

The process of adding a remote hash-lock to the gateway protocol
is summarized in Figure~\ref{fig:remote-hash-lock},
where a hash-lock smart contract HL4 is assumed to be located at a ``remote'' blockchain B4.
The hash-lock construct is assumed to be located at B4 because
under the opaque ledgers principle (Section~\ref{subsec:DesignPrinciples})
the blockchains B1 and B2 are inaccessible to external parties. 
Thus, even if B1 had smart-contract capabilities to implement a hash-lock,
it will only be readable (i.e. can be invoked) 
by entities in B1 only (and vice versa for blockchain B2).
We refer to such remote hash-locks as ``delegated'' hash-locks because
it is in essence a {\em delegated authorization} from the originator (Alice)
to the owner of gateway G1 (namely the VASP entity V1) to have G1 invoke the smart contract hash-lock on B4.
The blockchain B4 with ledger L4 is assumed to be read/write accessible
to G1 and G2, which operate under the control of VASP entities V1 and V2 respectively.

Delegated hash-locks may play an important role is assisting
in the movement of fungible asset from one private blockchain to another private blockchain.
In the example of Figure~\ref{fig:remote-hash-lock},
blockchain B4 may implement a stablecoin (or CBDC) denomination that is
acceptable to the VASP~V1 (owner of gateway G1)
and to VASP~V2 (owner of gateway G2).
The joint-execution by G1 and G2 of the gateway protocol at the mechanical plane
-- that moves the fungible asset {\em representation} from blockchains B1 to B2 -- 
must accompanied in parallel with the {\em transfer of value} at the value plane.
Thus, looking at Figure~\ref{fig:remote-hash-lock},
there is in fact two parallel flows occurring 
(one flow in the mechanical plane and one flow at the value plane).
The role of the gateway is thus to inter-link or ``unite''
these two flows so that consistency is achieved at both planes.
These two unidirectional flows must be synchronized in order
to achieve consistency in both ledger without any loss of value 
(other than perhaps some transaction fees).
The need for a stable and common value denomination is therefore crucial
not only in our current case of a unidirectional asset movement,
but it is also core to the
concept of bidirectional atomic swaps and deals~\cite{HerlihyLiskov2019}
where the mediating denomination must remain unchanged (non-fluctuating)
during the entire swap~\cite{LiptonTreccani2021-book}.
This is one reason why the Bitcoin cryptocurrency~\cite{Bitcoin,GriffinShams2019,Godbole2020-coindesk} 
cannot be a mediating denomination for atomic swaps\footnote{Putting it in another way,
{\em ``Bitcoin has no value, and hence it can have any price''}~\cite{LiptonTreccani2021-book}.}.
A similar limitation appears to be true also for ethers~\cite{Shevchenko2020,VoellFoxley2020-coindesk}
in the Ethereum platform~\cite{Buterin2014}.

The basic idea is for G1 to use the hash-lock HL4
as means to ensure that gateway G2 (i.e. VASP~V2) obtains payment in the form of stablecoins in B4
prior to gateway G2 regenerating the asset representation at blockchain B2.
Using hash-lock HL4,
the gateway G1 locks $X$ number of stablecoins to the address of G2 within blockchain B4.
This is shown as Step~(a) in Figure~\ref{fig:remote-hash-lock}.
The amount $X$ is the equivalent value of Alice's fungible asset in B1 being moved
to Bob at blockchain B2.
If gateway G2 fails to regenerate the asset representation at B2 (as visible to Bob)
before the time $t_h$ expires,
the stablecoins in B4 reverts back to G1 (i.e. VASP~V1).

During the joint execution of the gateway protocol (see Figure~\ref{fig:gateway-phases}),
the gateway G1 must indicate to gateway G2
that the stablecoins have been put-aside (hash-locked) for G2 at the remote blockchain B4.
This message can be included in Step~3 (escrow evidence) of Phase~2 of Figure~\ref{fig:gateway-phases},
which will motivate G2 to complete the execution of the gateway protocol.
Since G2 has access to ledger L4 at blockchain B4,
it can see (read) the complete smart contract HL4
and the fact that an instance of HL4 has been invoked
by G1 with the stablecoins addressed to G2 
(to be released upon input of the secret $s$).

Towards the end of Phase~3 of the gateway protocol execution,
the gateway G1 delivers the secret $s$ to gateway G2
to permit gateway G2 (VASP~V2) to receive the stablecoins at blockchain B4.
This is shown as Step~(b) and Step~(c) in Figure~\ref{fig:remote-hash-lock}.
Upon receiving the stablecoins at B4
the gateway G2 issue (regenerates) the virtual asset on ledger L2, 
assigned to Bob's public-key at B2. 
This release of the secret $s$
can be incorporated into the gateway protocol,
such as within Step~9 (commit final) of Figure~\ref{fig:gateway-phases}.
This permits gateway G2 to regenerate the asset in ledger L2
in Step~10 of Figure~\ref{fig:gateway-phases}.

It is important to note that there is always the chance that VASP~V2 (gateway G2) is dishonest
and purposely fails to regenerate the virtual asset in ledger L2 at blockchain B2,
despite having received payment of $X$ units of stablecoins via the hash-lock at
the remote blockchain B4.
In such a case,
gateway G1 can dispute this dishonest behavior by using the fact that
(a) the gateway G2 had issued the signed prepare-commit message in Step~7 of Phase~2 of Figure~\ref{fig:gateway-phases},
and (b) the evidence on ledger L4 in blockchain B4 that gateway G2
has received a payment of $X$ number of stablecoins.

\section{Conclusions}
\label{sec:Conclusions}

If blockchain technology is to be a foundational infrastructure 
for the future global virtual assets industry,
then the currently nascent industry needs to begin addressing the challenges
around blockchain interoperability.
In the current we have discussed the need for blockchain gateways
as a means to achieve interoperability across different blockchain networks,
and to support the cross-blockchain mobility of virtual assets.

One key component for interoperability is the well-designed architecture for gateway nodes
that interconnect the various blockchain networks,
based on standardized APIs and gateway protocols.
We discuss some of the design principles for gateways,
notably under the strict condition where the origin blockchain
and the destination blockchain are both private/permissioned.

Another key ingredient is a standard gateway-to-gateway protocol that implements
the movement of virtual assets from one blockchain to another
satisfying the classic ACID properties
(atomicity, consistency, isolation and durability).
We discuss an example of a gateway protocol
for the unidirectional movement of assets,
following the classic 2-Phase Commit paradigm.
The unidirectional protocol can be used as a building-block
to achieve ``atomic swaps'' across two private blockchain systems.

Finally,
we discuss several aspects of gateway deployments,
including gateway identities and certificates,
gateway crash recovery,
and the problem of the discoverability of gateways at a global scale.



\end{document}